\documentclass[runningheads]{llncs}
\usepackage{graphicx}
\usepackage{hyperref}
\begin{document}
\title{Mapping RDF Graphs to Property Graphs}
\author{Shota Matsumoto\inst{1} \and Ryota Yamanaka\inst{2} \and Hirokazu Chiba\inst{3}}
\authorrunning{S. Matsumoto et al.}
\institute{
Lifematics Inc., Tokyo 101-0041, Japan\\
\email{shota.matsumoto@lifematics.co.jp}
\and
Oracle Corporation, Bangkok 10500, Thailand\\
\email{ryota.yamanaka@oracle.com}
\and
Database Center for Life Science, Chiba 277-0871, Japan\\
\email{chiba@dbcls.rois.ac.jp}
}
\maketitle              % typeset the header of the contribution
\begin{abstract}
% The abstract should briefly summarize the contents of the paper in 15--250 words.
Increasing amounts of scientific and social data are published in the Resource Description Framework (RDF). Although the RDF data can be queried using the SPARQL language, even the SPARQL-based operation has a limitation in implementing traversal or analytical algorithms.
Recently, a variety of graph database implementations dedicated to analyses on the property graph model have emerged.
However, the RDF model and the property graph model are not interoperable.
Here, we developed a framework based on the Graph to Graph Mapping Language (G2GML) for mapping RDF graphs to property graphs to make the most of accumulated RDF data.
Using this framework, graph data described in the RDF model can be converted to the property graph model and can be loaded to several graph database engines for further analysis. Future works include implementing and utilizing graph algorithms to make the most of the accumulated data in various analytical engines.

\end{abstract}

\section{Introduction}
% Please note that the first paragraph of a section or subsection is not indented. The first paragraph that follows a table, figure, equation etc. does not need an indent, either.

Increasing amounts of scientific and social data are described as graphs. As a format of graph data, the Resource Description Framework (RDF) is widely used. Although RDF data can be queried using the SPARQL language in a flexible way, SPARQL is not dedicated to traversal of graphs and has a limitation in implementing graph analysis algorithms.

In the context of graph analysis, the property graph model~\cite{angles} is becoming popular; various graph database engines, including Neo4j~\cite{neo4j}, Oracle Labs PGX~\cite{pgx}, and Amazon Neptune~\cite{neptune}, adopt this model. These graph database engines support algorithms for traversal or analyzing graphs. However, currently not many datasets are consistently described in the property graph model, so the application of these powerful engines are limited.

Considering this situation, it is valuable to develop a method to transform RDF data into property graphs. However, the transformation is not straightforward due to the differences in the data model. 
In RDF graphs, all information is expressed as the triple (node-edge-node), whereas in property graphs, arbitrary information can be contained in each of the nodes and edges as key-value form. 
Although previous works addressed this issue by formalizing transformations~\cite{hartig},
users cannot define their specific mappings intended for each use case.

Here, we developed a framework based on the Graph to Graph Mapping Language (G2GML) for mapping RDF graphs to property graphs. Using this framework, accumulated graph data described in the RDF model can be converted to the property graph model and can be loaded to several graph database engines. 

\section{Methods}

\begin{figure}
\center
\includegraphics[width=1.0\textwidth]{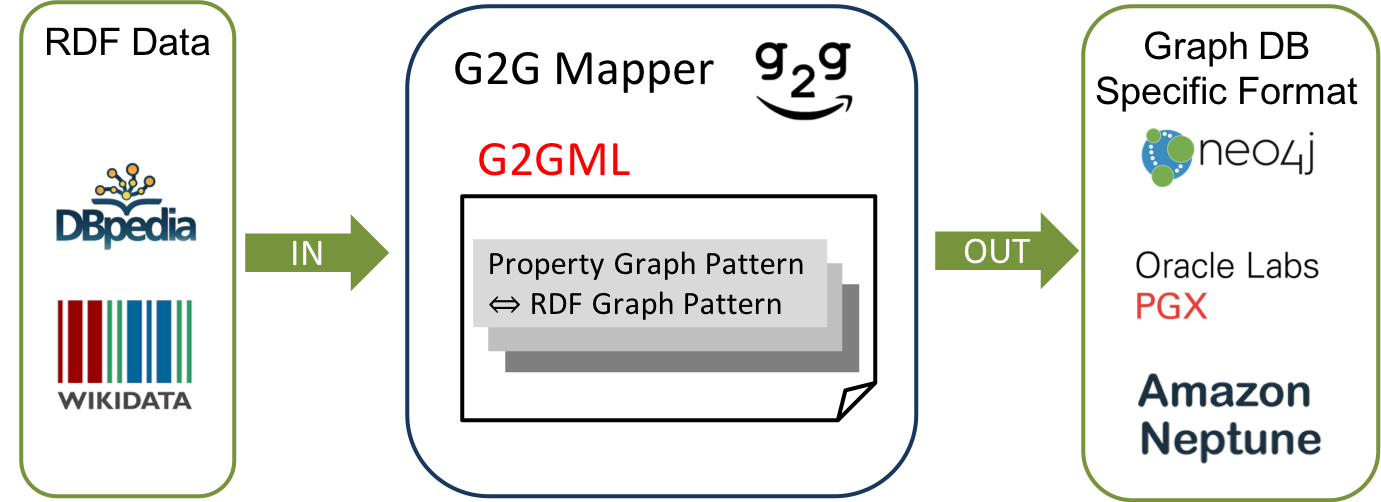}
\caption{Overview of G2GML mapping}
\label{fig:dataflow}
\end{figure}

Figure~\ref{fig:dataflow} shows the overview of proposed framework.
In the proposed framework, users write mappings from RDF graphs to property graphs in G2GML.
This mapping can be processed by an implementation called \textit{G2G Mapper}, which is implemented by authors (available on \url{https://github.com/g2gml}). This tool retrieves RDF data from SPARQL endpoints and converts them to property graph data in several different formats specified by popular graph databases.

G2GML is a declarative language which consists of pairs of RDF graph patterns and property graph patterns. 
An intuitive meaning of a G2GML is a mapping between RDF subgraphs that matches the described patterns and described components of the property graph. In the next section, we briefly explain the syntax of G2GML with a concrete example usage.

\section{Example}

\begin{figure}
\center
\includegraphics[width=1.0\textwidth]{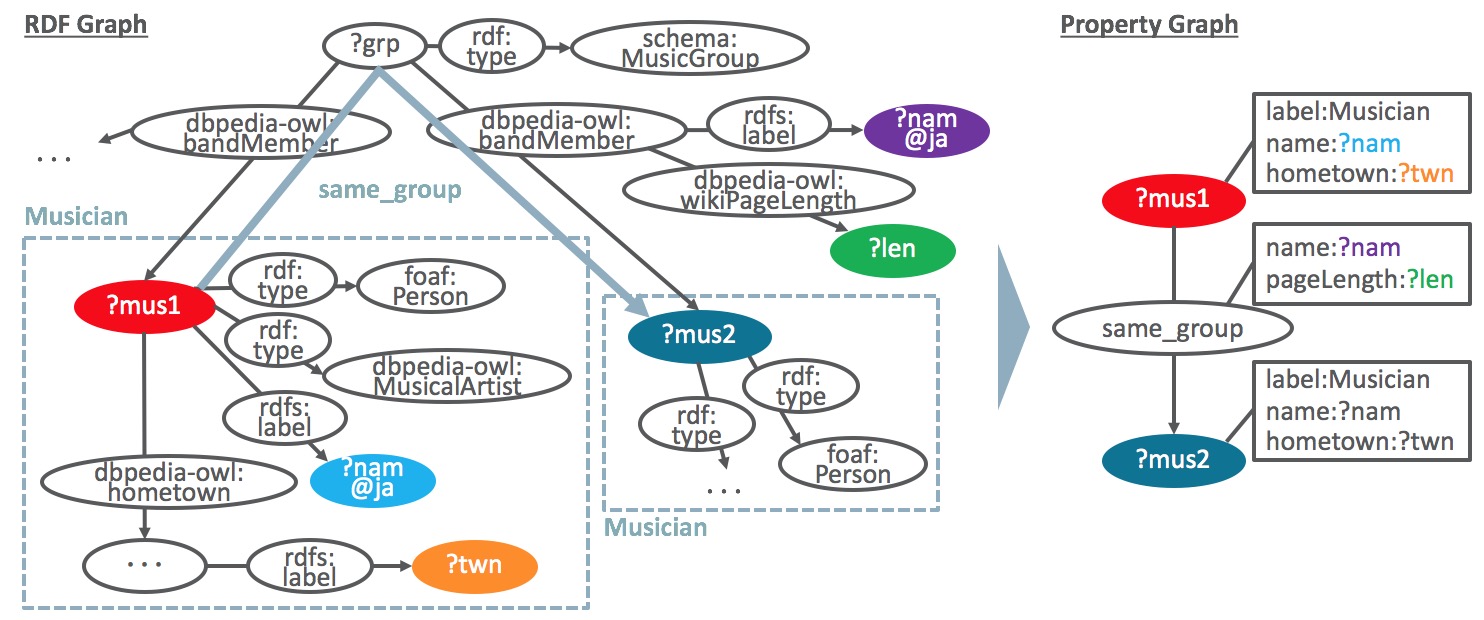}
\caption{Mapping of RDF data}
\label{fig:conversion}
\end{figure}

Figure~\ref{fig:conversion} shows an example of G2GML mapping, which converts RDF data retrieved from DBpedia into property graph data. When we focus on relationships that one musician and another are in the same group, the information can be summarized into the property graph data as shown in this figure.

For this conversion, the actual G2GML is described as in Figure~\ref{fig:g2gml}. It starts with URI prefixes used to write mappings, and then, each mapping consists of one unindented line of a property graph pattern and indented lines of an RDF graph pattern. A property graph pattern is written in a syntax like Cypher (the query language of Neo4j), whereas an RDF graph pattern is written as a pattern in SPARQL. Variables in each pattern are mapped by those names. This example contains one node mapping for \texttt{Musician} entity and one edge mapping for \texttt{same\_group} relationship only. In G2GML, edge mappings are defined based on the conditions of node mappings, which means that edges are generated in property graph iff both nodes' patterns and edges' patterns are matched in RDF graph. Also, \texttt{mus, nam, dat, twn} and \texttt{len} are used as variables to extract resources and literals from RDF graph. In the resulting property graph, resources can be mapped to nodes, while literals can be mapped to values of properties.

\begin{figure}[!t]
\vspace{2mm}
\begin{scriptsize}
\begin{verbatim}
# Prefixes
PREFIX rdf: <http://www.w3.org/1999/02/22-rdf-syntax-ns#>
PREFIX rdfs: <http://www.w3.org/2000/01/rdf-schema#>
PREFIX prop: <http://dbpedia.org/property/>
PREFIX schema: <http://schema.org/>
PREFIX dbpedia-owl: <http://dbpedia.org/ontology/>
PREFIX foaf: <http://xmlns.com/foaf/0.1/>

# Node mapping
(mus:Musician {vis_label:nam, born:dat, hometown:twn})                    # PG Pattern
    ?mus rdf:type foaf:Person, dbpedia-owl:MusicalArtist .                # RDF Pattern
    ?mus rdfs:label ?nam .
    OPTIONAL { ?mus prop:born ?dat }
    OPTIONAL { ?mus dbpedia-owl:hometown / rdfs:label ?twn }

# Edge mapping
(mus1:Musician)-[:same_group {label:nam, length:len}]->(mus2:Musician)    # PG Pattern
    ?grp a schema:MusicGroup ;                                            # RDF Pattern
         dbpedia-owl:bandMember ?mus1 , ?mus2 .
    FILTER(?mus1 != ?mus2)
    OPTIONAL { ?grp rdfs:label ?nam. FILTER(lang(?nam) = "ja")}
    OPTIONAL { ?grp dbpedia-owl:wikiPageLength ?len }
\end{verbatim}
\end{scriptsize}
\caption{G2GML mapping definition}
\label{fig:g2gml}
\end{figure}

Finally, Figure~\ref{fig:sparql} shows the SPARQL query to retrieve the pairs of musicians who are in the same group. After G2GML mapping above, we can load the generated property graph data into graph databases, such as Oracle Labs PGX, and the query can be written in PGQL (the query language of PGX).

\begin{figure}[!t]
\vspace{2mm}
\begin{scriptsize}
\begin{verbatim}
# SPARQL
PREFIX rdf: <http://www.w3.org/1999/02/22-rdf-syntax-ns#>
PREFIX rdfs: <http://www.w3.org/2000/01/rdf-schema#>
PREFIX schema: <http://schema.org/>
PREFIX dbpedia-owl: <http://dbpedia.org/ontology/>
SELECT DISTINCT
    ?nam1 ?nam2
WHERE {
    ?mus1 rdf:type foaf:Person , dbpedia-owl:MusicalArtist .
    ?mus2 rdf:type foaf:Person , dbpedia-owl:MusicalArtist .
    ?mus1 rdfs:label ?nam1 . FILTER(lang(?nam1) = "ja") .
    ?mus1 rdfs:label ?nam2 . FILTER(lang(?nam2) = "ja") .
    ?grp a schema:MusicGroup ;
         dbpedia-owl:bandMember ?mus1 , ?mus2 .
    FILTER(?mus1 != ?mus2)
}

# PGQL
SELECT DISTINCT m1.name, m2.name WHERE (m1)-[same_group]-(m2)
\end{verbatim}
\end{scriptsize}
\caption{SPARQL and PGQL}
\label{fig:sparql}
\end{figure}

\section{Conclusion}
In this work, we defined G2GML for mapping RDF graphs to property graphs and implemented a converter based on the G2GML. We also showed an example usage of G2GML. Future works include further analysis of the converted graph data on the database engines adopting the property graph model.


\end{document}